\documentclass[journal]{vgtc}       

\AtEndPreamble{\hypersetup{linkcolor=black,citecolor=black,urlcolor=black}}
\AtEndPreamble{\crefname{figure}{Figure}{Figures}\Crefname{figure}{Figure}{Figures}\crefname{table}{Table}{Tables}}

\AtBeginDocument{\setlength{\tabcolsep}{3pt}}

\onlineid{0}

\vgtccategory{Research}

\title{3D Field Data Reduction with Adaptive Sample-Based Gaussian-Encoded Reconstruction}

\author{Michael R. Martin, Joseph Insley, Victor A. Mateevitsi, Silvio Rizzi, and Kwan-Liu Ma}

\authorfooter{
  \item Michael R. Martin and Kwan-Liu Ma are with the Department of Computer Science, University of California, Davis.
        E-mail: csemartin@ucdavis.edu, klma@ucdavis.edu.
  \item Joseph Insley, Victor A. Mateevitsi, and Silvio Rizzi are with Argonne National Laboratory.
        E-mail: insley@anl.gov, vmateevitsi@anl.gov, srizzi@anl.gov.
}

\abstract{
In scientific simulation, regular grids, unstructured meshes, and particle-based formats are chosen to represent field data for computational efficiency, geometry/adaptive flexibility, and following motion/deformation, respectively. Each of these field data formats is often handled through separate data-specific processing pipelines. We present a unified sample-based Gaussian encoding method that represents these data forms under a single fixed-budget formulation. The method initializes and refines Gaussian primitives directly from the input samples while preserving a prescribed primitive count and encoded size to achieve a desired level of data reduction. Across structured, unstructured, and particle data, the sample-based formulation improves reconstruction accuracy with measurably fewer primitives in comparison to prior formulations, achieving up to 4.8~dB higher PSNR with an approximate $44\times$ reduction in primitive count. For time-varying data, warm-starting from the previous timestep reduces the optimization required to reach independently trained reconstruction quality. Together, these results demonstrate a unified fixed-budget Gaussian encoding framework for structured, particle, unstructured, and time-varying scientific data with predictable storage, higher reconstruction accuracy, and improved temporal encoding efficiency.
}


\usepackage{mathptmx}
\usepackage{booktabs}
\usepackage{array}
\usepackage{multirow}
\usepackage{amsmath}
\usepackage{comment} 

\graphicspath{{figures/}}

\begin{document}
\raggedbottom
\firstsection{Introduction}
\maketitle
\label{sec:intro}
Scientific simulations represent scalar fields in forms required by their numerical methods. Cosmological hydrodynamics codes use particles with smoothing lengths, allowing resolution to follow moving and deforming matter~\cite{habib2014hacc}. Hydrocodes for shocks and impacts use adaptive unstructured meshes whose cells concentrate resolution near steep gradients~\cite{gittings2008rage}. Other solvers use regular grids for their fixed stencils and uniform I/O. To visualize particle and mesh data, pipelines commonly deposit or resample onto a regular grid, which establishes a uniform-resolution representation for the entire domain, and can create dense data products whose storage and movement grow with simulation resolution. For time series, these costs recur at every stored step; at HPC scale, full-resolution fields must be written, transferred, and loaded before post hoc analysis can begin~\cite{ma2009insitu,yu2010insitu}. Data-reduction methods address parts of this problem, but their assumptions remain tied to specific data forms or representation choices. Error-bounded compressors operate on voxel arrays and reconstruct them upon decompression~\cite{lindstrom2014zfp,di2016sz}. Implicit neural representations (INRs) can provide high field fidelity per byte, but their storage and query cost are determined by a selected neural-network architecture and, where used, the learned feature-table configuration rather than an explicit count of renderable elements~\cite{lu2021compressive,muller2022instant,weiss2022fast}. While Gaussian primitive formulations support flexible post hoc rendering~\cite{dyken2025veg,dyken2026wveg}, existing pipelines either adjust capacity dynamically through optimization-time densification~\cite{dyken2026wveg}, rely on image-space supervision~\cite{jiang2026particlegs}, or restrict fixed-budget primitive allocation strictly to regular-grid inputs~\cite{martin2026fixedbudget}. This leaves a clear gap: a scalar-field representation with a predictable, pre-allocated memory footprint whose capacity allocation operates independently of grid structures. We address this gap with a sample-based fixed-budget Gaussian encoder that provides one representation for particles, unstructured cells, and regular-grid scalar fields. Rather than requiring a regular grid for capacity allocation, the method constructs a prescribed set of scalar-carrying anisotropic Gaussians from the source data. The fixed primitive budget determines the full-precision model footprint before training, while the retained scalar attributes allow one encoding to support post hoc transfer-function, lighting, and viewpoint changes. The formulation also extends to time-varying fields through temporal reuse. Our goal is to evaluate whether this unified fixed-budget formulation can compress structured grids, unstructured cells, particles, and time-varying fields with predictable storage while preserving reconstruction quality. We make four contributions: \textbf{(i)} a sample-based Gaussian encoding formulation that accepts particle, unstructured-cell, and regular-grid scalar data under one fixed primitive budget; \textbf{(ii)} a budget-preserving refinement strategy that improves the use of the prescribed representation capacity without changing its encoded size; \textbf{(iii)} a temporal warm-start strategy and empirical evaluation of the conditions under which it approaches independently trained reconstruction quality; and \textbf{(iv)} an evaluation across particle, structured-grid, unstructured-cell, and time-varying data, including controlled matched-budget comparisons with the grid-seeded predecessor; near-matched-count comparisons against a learned-Gaussian encoder, evaluated under its public release schedule; and size-fidelity reference points from implicit neural representations.
\section{Related Work}
\label{sec:related}

\subsection{Gaussian representations for scalar volumes.}
3D Gaussian Splatting represents scenes with anisotropic Gaussian primitives optimized by differentiable image rendering~\cite{kerbl2023gaussian}. Its density-control variants adapt the primitive set during optimization through splitting, insertion, or pruning~\cite{rotabulo2024revising,kheradmand2024mcmc}. Volume Encoding Gaussians (VEG) adapt Gaussian primitives to scalar volumes, retaining scalar attributes so that transfer functions can be applied at render time~\cite{dyken2025veg}. VEG is trained from rendered images, whereas W-VEG supervises the reconstructed scalar field at sampled three-dimensional positions and supports structured and unstructured volumes~\cite{dyken2026wveg}. W-VEG allocates capacity through error-guided densification and pruning toward a compression-derived target. Martin et al.~\cite{martin2026fixedbudget} developed a fixed-budget Gaussian encoder for regular-grid scalar volumes allocating a complete scalar-carrying anisotropic primitive set before field-space refinement, making the full-precision storage footprint a known specification rather than an optimization outcome. This representation retains the transfer-function flexibility of scalar-field encoding, with its allocation and supervision centered on regular-grid data. Our sample-based formulation preserves this prescribed-capacity scalar representation while extending the method to native particle and unstructured-cell inputs, budget-preserving primitive relocation, and temporal reuse. iVR-GS~\cite{tang2025ivrgs} and ECoNGS~\cite{tang2026econgs} investigate editable and compressed Gaussian representations for volume visualization. Complementary methods reduce the storage of an already trained Gaussian model~\cite{fan2024lightgaussian,niedermayr2024compressed,papantonakis2024reducing,chen2024hac}, whereas OpenVDB-oriented methods ~\cite{sharma2025vdbgaussian,sharma2025vdbraymarch} approximate hierarchical sparse volumes with Gaussian-based representations and ray marching. 
\subsection{Neural volume representations.} 
Coordinate networks represent scalar fields as continuous functions of position~\cite{sitzmann2020siren,lu2021compressive}, while feature-grid and multiresolution hash encodings combine learned feature tables with a small decoder network~\cite{muller2022instant,wu2024interactive,weiss2022fast}. NeuralVDB~\cite{kim2024neuralvdb} and adaptively placed multi-grid networks~\cite{wurster2024adaptively} extend neural representations to large and spatially heterogeneous volumes. UGINR fits an implicit representation directly to unstructured grids~\cite{liu2024uginr}. These methods can provide compact, high-fidelity field approximations, but their storage footprint is governed by architectural and feature-table choices, and field evaluation requires network inference rather than querying an explicit primitive set. 
\subsection{Time-varying fields and conventional compression.} 
Time-varying volume visualization is constrained by storage, movement, and resident-memory costs of successive fields~\cite{ma2003timevarying}. MoE-INR represents time-varying volumes with a mixture-of-experts implicit model~\cite{han2026moeinr}, and ANTIC adapts neural compression to temporal in situ data~\cite{antic2026}. Conventional compressors operate directly on voxel arrays: adaptive prediction-based methods provide lossless floating-point compression~\cite{fout2012lossless}, while zfp ~\cite{lindstrom2014zfp} and SZ ~\cite{di2016sz} provide fixed-rate or error-bounded lossy compression. These approaches reconstruct voxel arrays and are complementary to learned representations storing a field model. 
\subsection{Particle and unstructured grid data.} 
Large unstructured-grid volume rendering remains challenging because the representation must retain mesh connectivity and spatially varying cell geometry~\cite{sarton2023sota}. Memory-efficient encodings and GPU-oriented partitioning improve direct rendering while preserving the mesh~\cite{wald2022memory,morrical2023quick}. ParticleGS applies Gaussian splatting to HACC particle data and optimizes image fidelity for selected visualization settings~\cite{jiang2026particlegs}. These methods establish representations for individual data forms but they do not provide a single fixed-budget, scalar-field encoding formulation across particles, unstructured cells, and regular-grid samples.
\section{Method}\label{sec:method}

The sample-based encoder preserves the scalar-carrying Gaussian representation, fixed primitive budget, and export format of the grid-seeded encoder~\cite{martin2026fixedbudget}. It replaces the grid-specific allocation and refinement stages with a common sample formulation. Particles, unstructured cells, and grid voxels provide samples for initialization; all models are then refined against a field-space reference. Fixed-count relocation reallocates underused capacity without changing the encoded size, and a warm start reuses a model across stored time steps.

\subsection{Representation and Export}\label{sec:representation}

A model contains $N$ anisotropic Gaussian primitives. Primitive $i$ stores a center $\boldsymbol{\mu}_i$, log-scales $\boldsymbol{\ell}_i$, a quaternion $\mathbf{q}_i$, a scalar logit $\alpha_i$, and a weight logit $\beta_i$. Its scalar value and weight are
\begin{equation}
s_i=\operatorname{sigmoid}(\alpha_i), \qquad w_i=\operatorname{sigmoid}(\beta_i).
\end{equation}
The quaternion defines the rotation $\mathbf{R}_i$, and the standard deviations used in field evaluation are $\boldsymbol{\sigma}_i=\operatorname{clamp}(\exp(\boldsymbol{\ell}_i),0.4,8.0)$ in voxel units. Let $d_i^2(\mathbf{x})$ denote the squared, rotated, scale-normalized distance to primitive $i$. The truncated kernel is
\begin{equation}
\begin{aligned}
d_i^2(\mathbf{x}) &= \left\|\operatorname{diag}(\boldsymbol{\sigma}_i)^{-1}\mathbf{R}_i^{\mathsf{T}}(\mathbf{x}-\boldsymbol{\mu}_i)\right\|_2^2, \\
K_i(\mathbf{x}) &= \exp\!\left(-\tfrac{1}{2}d_i^2(\mathbf{x})\right)\mathbf{1}[d_i^2(\mathbf{x})<25],
\end{aligned}
\label{eq:kernel}
\end{equation}
which truncates each Gaussian at five standard deviations. Here, $\mathbf{1}[\cdot]$ is an indicator function that equals one when its
condition holds and zero otherwise. The mixture density and scalar numerator are
\begin{equation}
\rho(\mathbf{x})=\sum_{i=1}^{N}w_iK_i(\mathbf{x}), \qquad n(\mathbf{x})=\sum_{i=1}^{N}w_iK_i(\mathbf{x})s_i.
\label{eq:density_numerator}
\end{equation}
Training uses $\widehat{V}_{\mathrm{train}}(\mathbf{x})=n(\mathbf{x})/\max(\rho(\mathbf{x}),10^{-3})$ as a numerical denominator guard. Every decoded grid, rendered image, and reported voxel metric instead uses
\begin{equation}
\widehat{V}_{\mathrm{eval}}(\mathbf{x})=\operatorname{clamp}_{[0,1]}\left(\frac{n(\mathbf{x})}{\max(\rho(\mathbf{x}),0.05)}\right).
\label{eq:decoded_field}
\end{equation}
The coverage objective of Section~\ref{sec:objective} promotes $\rho(\mathbf{x})\geq0.05$ at sampled locations where $V_{\mathrm{ref}}(\mathbf{x})>0.005$. Thus, the training and evaluation definitions coincide where this coverage condition is satisfied, while the larger evaluation floor suppresses unreliable values from distant Gaussian tails in empty space. As budget $N$ is fixed pre-training, the full-precision export stores twelve float32 parameters per primitive (three center coordinates, three log-scales, four quaternion components, one scalar logit, and one weight logit) for 48 bytes per primitive. The mixed-precision export stores center coordinates as float16 and quantizes the remaining nine parameters to unit8 using per-array min-max affine quantization, yielding 15 bytes of parameter payload per primitive. Lossless compression of this archive yields approximately 12 bytes per primitive on disk. The representation remains scalar, so one model can be decoded at a chosen resolution and rendered under different transfer functions, lighting, and viewpoints.

\begin{table}[h]
  \raggedright
  \vspace{-2.5pt}
  \caption{Sample definitions for each input form. All samples are expressed in the index frame of the reference grid; $hh$ denotes the SPH smoothing length, and $\gamma$ converts Deep Water Impact source-space lengths to reference-grid index units.}
  \vspace{-5pt} 
  \label{tab:samples}
  \scriptsize
  \setlength{\tabcolsep}{2pt}
  \begin{tabular}{@{}p{0.16\linewidth}p{0.11\linewidth}p{0.13\linewidth}p{0.18\linewidth}p{0.10\linewidth}p{0.23\linewidth}@{}}
    \toprule
    Input form & $\mathbf{q}_j$ & $h_j$ & $v_j$ & $m_j$ & Reference field $V_{\mathrm{ref}}$ \\
    \midrule
    HACC SPH particles & particle position & $0.8\,hh$ & normalized log density & particle mass & particle deposit \\
    Deep Water Impact cells & cell center & $\tfrac{1}{2}\gamma\,\mathrm{vol}^{1/3}$ & normalized cell value & cell volume & volume-weighted cell deposit \\
    Grid voxels & voxel center & $\tfrac{1}{2}$ voxel & normalized voxel value & 1 & input grid \\
    \bottomrule
  \end{tabular}
  \vspace{-10pt} 
\end{table}

\subsection{Sample Interface}\label{sec:samples}

The encoder consumes a set of samples $\{(\mathbf{q}_j,h_j,v_j,m_j)\}_{j=1}^{N_s}$, where $\mathbf{q}_j$ is position, $h_j$ is an initial local scale, $v_j$ is the normalized scalar value, and $m_j$ is an importance weight. Positions and scales are expressed in the index frame of a chosen reference grid, with one voxel as the unit of length. Values are normalized to $[0,1]$ using the field range; one range is fixed across a time series. Table~\ref{tab:samples} defines the samples for each input form. For HACC data \cite{habib2014hacc}, input samples are SPH gas particles: after conversion to the reference-grid index frame, $h_j=0.8\,hh_j$, where $hh_j$ is the particle smoothing length; $v_j$ is normalized log density and $m_j$ is particle mass. For regular grids, each voxel is a sample with half-voxel initial scale and unit importance. For Deep Water Impact data \cite{patchett2017deepwater}, the input samples are the centers of axis-aligned hexahedral cells. Let $\gamma=\min_a D_a/(b_a^{\max}-b_a^{\min})$ convert source lengths to the index frame of a reference grid with dimensions $\mathbf{D}$ and bounds $[\mathbf{b}^{\min},\mathbf{b}^{\max}]$. A cell of source-space volume $\mathrm{vol}_j$ receives $h_j=\tfrac{1}{2}\gamma\,\mathrm{vol}_j^{1/3}$, and its importance weight is $\mathrm{vol}_j$. Particle and cell samples initialize the Gaussian model directly. A reference grid supplies field-space supervision and evaluation rather than primitive seeds: HACC uses a particle deposit, whereas grid data use the input grid itself. For unstructured cells, we construct a volume-weighted deposited reference. Cell $j$ stamps the clamped voxel-index box $[\lfloor\mathbf{q}_j-2h_j\rfloor,\lceil\mathbf{q}_j+2h_j\rceil)$, and each covered voxel stores
\begin{equation}
V_{\mathrm{ref}}(\mathbf{u})=\frac{\sum_{j\in\mathcal{C}(\mathbf{u})}m_jv_j}{\sum_{j\in\mathcal{C}(\mathbf{u})}m_j},
\label{eq:cell_deposit}
\end{equation}
with value zero when no cell stamp contains $\mathbf{u}$. Here, $\mathcal{C}(\mathbf{u})$ denotes the set of cell stamps containing voxel $\mathbf{u}$; membership is defined by the preceding stamped index box rather than by exact cell--voxel overlap volumes. Hence, the deposited grid is a common field-space target, while initialization remains derived from native particle or cell samples.

\subsection{Seeding from Samples}\label{sec:seeding}

The prescribed budget $N$ is allocated through three sample draws. An importance stream receives 45\% of the budget and samples without replacement with probability proportional to $m_j^{0.6}$. The exponent tempers the dynamic range of the importance weights, preventing a few highly weighted samples from consuming the complete budget. A uniform stream receives 40\% of the budget and provides coverage across the sample set. The remaining 15\% form a peak stream: within each $2{\times}2{\times}2$-voxel spatial bin, it retains the highest-valued sample when that value exceeds 0.55. If fewer bins qualify, their unused allocation is reassigned to the uniform stream. For the importance and uniform streams, let $N_s$ denote the number of input
samples and $N$ the prescribed primitive budget. Each selected sample initializes one isotropic, unrotated Gaussian with its sample position and scalar value. Importance and uniform seeds use
\begin{equation}
h_j\max\!\left(\left(\frac{N_s}{N}\right)^{1/3},1\right),
\label{eq:seed_scale}
\end{equation}

which widens their initial support according to the mean sample-spacing ratio. Peak seeds retain their local sample scale to preserve compact high-value structure. Initial scales are clipped to 0.35--6 voxels for the importance and uniform streams and 0.35--2.5 voxels for the peak stream, with initial weights of 0.35 and 0.6. For regular-grid input, $m_j=1$, so the importance stream becomes a second uniform draw.
\subsection{Fidelity Objective and Relocation}\label{sec:objective}

Let $\mathcal{P}$ denote the field-space points sampled at each refinement iteration and $\omega_p=\mathbf{1}[V_{\mathrm{ref}}(\mathbf{x}_p)>0.005](0.5+V_{\mathrm{ref}}(\mathbf{x}_p))$. Here, $p$ indexes a sampled point, and $\mathbf{1}[\cdot]$ has the same
indicator-function meaning as in Eq.~\eqref{eq:kernel}. In the scale terms below, $i$ indexes a Gaussian primitive and $a\in\{x,y,z\}$ indexes one of its three scale axes. We minimize
\begin{equation}
\mathcal{L}=\mathcal{L}_{\mathrm{val}}+2\mathcal{L}_{\mathrm{cov}}+0.1\mathcal{L}_{\mathrm{scale}}+0.05\mathcal{L}_{\mathrm{aniso}},
\label{eq:loss}
\end{equation}
where
\begin{align}
\mathcal{L}_{\mathrm{val}}&=\frac{\sum_{p\in\mathcal{P}}\omega_p\,\operatorname{Huber}_{0.1}\!\left(\widehat{V}_{\mathrm{train}}(\mathbf{x}_p)-V_{\mathrm{ref}}(\mathbf{x}_p)\right)}{\max(\sum_{p\in\mathcal{P}}\omega_p,10^{-6})}, \nonumber\\
\mathcal{L}_{\mathrm{cov}}&=\frac{1}{|\mathcal{P}|}\sum_{p\in\mathcal{P}}\mathbf{1}[V_{\mathrm{ref}}(\mathbf{x}_p)>0.005]\,\max(0,0.05-\rho(\mathbf{x}_p)), \nonumber\\
\mathcal{L}_{\mathrm{scale}}&=\underset{i,a}{\operatorname{mean}}\,\max(0,\exp(\ell_{i,a})-8) \nonumber\\
&\quad+0.05\,\underset{i,a}{\operatorname{mean}}\,\max(0,0.30-\exp(\ell_{i,a})), \nonumber\\
\mathcal{L}_{\mathrm{aniso}}&=\underset{i}{\operatorname{mean}}\,\max\!\left(0,\frac{\max_a\exp(\ell_{i,a})}{\max(\min_a\exp(\ell_{i,a}),10^{-3})}-4\right).
\label{eq:loss_terms}
\end{align}

Here, $\operatorname{Huber}_{0.1}$ denotes the Huber loss~\cite{huber1964robust} with transition parameter $0.1$, applied to the scalar reconstruction residual in $\mathcal{L}_{\mathrm{val}}$. At each refinement iteration, the sampled points are drawn from four strata: 30\% from high-valued locations ($V_{\mathrm{ref}}>0.4$), 30\% from mid-valued locations ($0.02<V_{\mathrm{ref}}\leq0.4$), 15\% uniformly over the domain, and 25\% from sample positions perturbed by Gaussian noise with standard deviation 0.75 voxel. Every point receives an additional half-voxel uniform jitter and is clamped to the domain. The reference value at each off-grid point is obtained by trilinear interpolation of $V_{\mathrm{ref}}$. Samples whose interpolated reference value is at most 0.005 are masked from the fidelity and coverage terms; the uniform stratum therefore supplies occupied-field samples outside the high- and mid-value pools. The fidelity term emphasizes high-valued structure, while the coverage term promotes sufficient mixture support at sampled locations where $V_{\mathrm{ref}}(\mathbf{x})>0.005$, without prescribing density elsewhere. The soft scale penalty bounds stored scales to a 0.30--8 voxel band, whereas kernel evaluation independently clamps active scales to 0.4--8 voxels. No density-uniformity or inter-primitive smoothness objective is used; the mixture density can therefore adapt to sharp, spatially localized structure. At 35\%, 55\%, and 75\% of refinement, a relocation pass evaluates a fresh stratified set of $2^{16}$ points. It selects the 8\% of primitives with the smallest effective support, $w_i\prod_a\exp(\ell_{i,a})$, and pairs them with the equal number of largest masked reconstruction errors
\begin{equation}
\left|\widehat{V}_{\mathrm{train}}(\mathbf{x})-V_{\mathrm{ref}}(\mathbf{x})\right|\mathbf{1}[V_{\mathrm{ref}}(\mathbf{x})>0.02].
\end{equation}
The selected primitives receive the corresponding error-point centers, an isotropic 0.8-voxel scale, the local reference scalar value, and weight 0.5. Relocation preserves both the primitive count and model footprint; subsequent iterations refine the relocated and unchanged primitives jointly.

\begin{figure*}[t]
\centering
\includegraphics[width=\textwidth,alt={Six HACC gas-density renderings: reference, 300K grid-seeded encoder, NeurComp, hash grid, and 300K 8-bit sample-based encodings supervised on 1024-cubed and 256-cubed deposits. Grayscale maps below each reconstruction show absolute luminance difference from the reference. The grid-seeded result has the broadest filaments and the most spatially distributed residual; both sample-based results retain separate small halos, while the hash grid has the smallest residual.}]{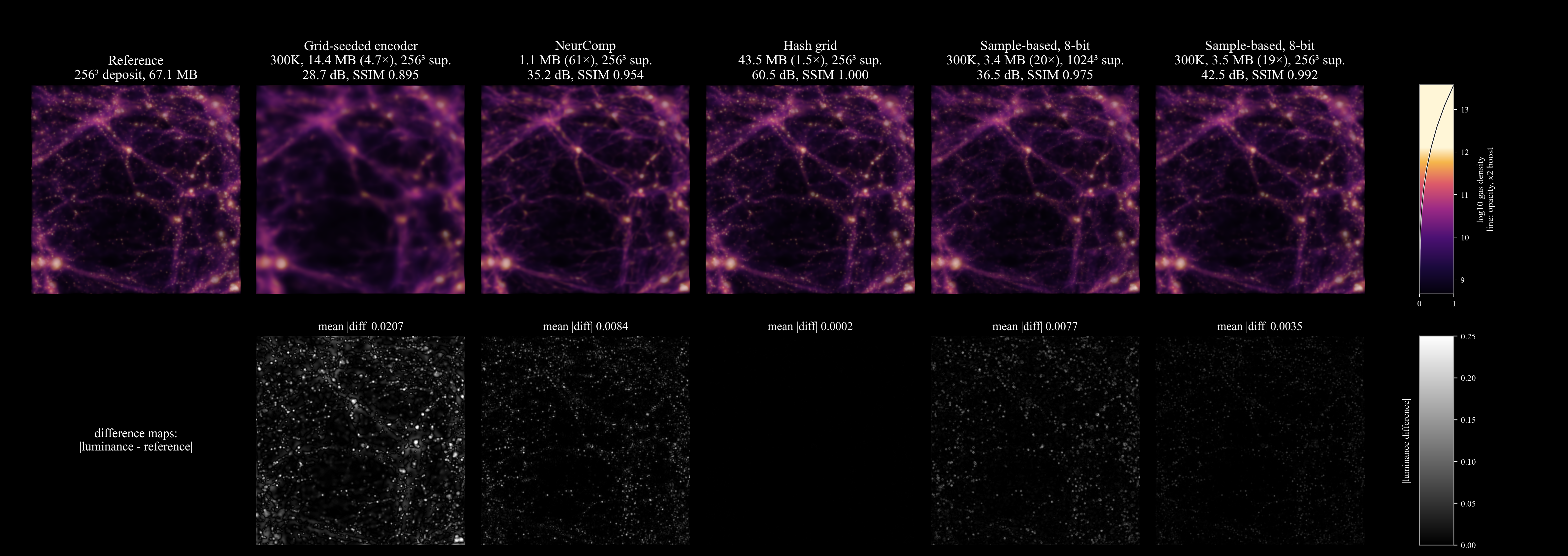}
\vspace{-12.5pt}
\caption{HACC rank-0 gas density on the $256^3$ reference deposit under a common camera and transfer function. Columns show reference, 300K grid-seeded encoder, implicit neural baselines, and 300K 8-bit sample-based exports with $1024^3$ and $256^3$ deposit supervision. Labels report on-disk size, reduction factor, supervision lattice, rendered PSNR, and SSIM. Bottom: absolute luminance difference from reference (black, 0; white, 0.25).}
\label{fig:hacc}
\vspace{-10pt} 
\end{figure*}

\subsection{Warm Start Across Timesteps}\label{sec:warmstart}

For a stored time series, the model for step $t$ initializes the model for step $t+1$. The center, log-scale, quaternion, scalar-logit, and weight-logit arrays are loaded directly; seeding is skipped and the loaded primitive count remains fixed. All steps use one normalization range for the series, after which the objective of Section~\ref{sec:objective} is applied unchanged. We evaluate two warm-start regimes: the coarse-stride regime refines the loaded arrangement without relocation, whereas the dense-stride regime retains the relocation passes so that capacity released by disappearing structure can be reassigned to newly formed structure. Each time step is stored as a complete model rather than as a delta from its predecessor. Relocation therefore improves temporal encoding efficiency without preserving a one-to-one correspondence between primitives across successive steps.
\section{Experimental Setup}
\label{sec:setup}

\subsection{Datasets}
\label{sec:datasets}

Table~\ref{tab:datasets} lists the six inputs and their evaluation references. HACC~\cite{habib2014hacc} contributes the SPH gas particles of one rank of a cosmology run. The primary HACC models use a $1024^3$ particle deposit for supervision and a $256^3$ particle deposit for evaluation; Table~3 additionally reports matched $256^3$-supervision controls. The unstructured-cell experiment uses Deep Water Impact~\cite{patchett2017deepwater}, an asteroid ocean-impact simulation: cycle 12{,}244 contributes 44,851,632  adaptive-mesh hexahedral cells. Separately, the temporal experiment uses 26 stored water-fraction fields from LANL’s $300^3$ regular-grid Deep Water Impact product, whose identical extent, origin, spacing, and dimensions provide a fixed lattice for warm-start evaluation; at cycle 17,260, we additionally encode the temperature and asteroid-fraction fields. Vortex, Bubble Plume, and Miranda~\cite{klacansky2017open} are regular-grid inputs: Vortex and Bubble Plume~\cite{martin2026fixedbudget} retain the grids used to evaluate the grid-seeded encoder, allowing its published models to be rescored. Miranda is evaluated at stride 3 ($342^3$). For the W-VEG baseline~\cite{dyken2026wveg}, the Miranda grid is converted to the explicit hexahedral mesh required by its public preprocessing pipeline; stride 3 is the finest converted mesh that completed in our 32\,GB host-memory environment.

\begin{table}[h]
\centering
\caption{Datasets. The input form is consumed by the sample-based encoder; the reference is the lattice on which reconstructed fields are evaluated. Grid sizes give float32 storage of the evaluated lattice.}
\vspace{-5pt} 
\label{tab:datasets}
\scriptsize
\setlength{\tabcolsep}{2.5pt}
\begin{tabular}{@{}>{\raggedright\arraybackslash}p{0.30\linewidth}>{\raggedright\arraybackslash}p{0.37\linewidth}>{\raggedright\arraybackslash}p{0.27\linewidth}@{}}
\toprule
Dataset & Input form and extent & Evaluation reference \\
\midrule
HACC gas, rank 0~\cite{habib2014hacc} & 1{,}395{,}369 SPH particles, $32\,h^{-1}$\,Mpc box & particle deposits, $1024^3$ supervision / $256^3$ evaluation \\
Deep Water Impact cells, cycle 12{,}244~\cite{patchett2017deepwater} & 44{,}851{,}632 hexahedral AMR cells, $46{\times}28{\times}24$\,km & volume-weighted cell deposit, $1024{\times}624{\times}536$ \\
Deep Water Impact temporal series & 26 LANL-distributed $300^3$ water-fraction grids & input grid (warm-start evaluation) \\
Vortex~\cite{martin2026fixedbudget} & $128^3$ grid, 8.4\,MB & input grid \\
Bubble, stride 2~\cite{martin2026fixedbudget} & $128{\times}128{\times}320$ grid, 21.0\,MB & input grid \\
Miranda, stride 3~\cite{klacansky2017open} & $342^3$ grid, 160\,MB & input grid \\
\bottomrule
\end{tabular}
\vspace{-10pt} 
\end{table}

\subsection{Training Configuration}
\label{sec:training}

All sample-based runs use Adam~\cite{kingma2015adam} with a learning rate of $8\times10^{-3}$ and cosine decay to 5\% of that rate. Models initialized independently use the seeding strategy of Section \ref{sec:seeding}, whereas warm-started steps load the preceding timestep’s model as described in Section \ref{sec:warmstart}. HACC, grid, and cell runs, as well as dense-stride warm starts, use three relocation passes; independently encoded time steps and coarse-stride warm starts omit relocation. The default full schedule draws 16{,}384 stratified points per iteration for 4{,}000 iterations on HACC and 3{,}000 iterations on grids, cells, and independently encoded time steps; the Miranda full-schedule run draws 8{,}192 points per iteration. The matched training-budget protocol uses the grid-seeded encoder's published iteration count and sample count, 1{,}500 iterations and 3{,}072 points per iteration. Each encoder retains its own initialization and refinement objective. For HACC, the grid-seeded run uses 300 iterations of this configuration. Short schedules draw 8{,}192 points per iteration, and warm-started steps use stride-dependent iteration counts. We abbreviate primitive budgets, for example, 32K for 32{,}000 primitives. Reported times measure only the optimization loop. They exclude sample-set construction, seeding, particle or cell deposition, mesh conversion, decoding, and export; W-VEG times similarly measure its training program after precomputed-sample generation. Experiments ran on an Alienware~16 Area-51 laptop with an Intel Core Ultra~9 275HX CPU, 32\,GB RAM, and an NVIDIA GeForce RTX~5090 Laptop GPU (driver 592.02), under Windows~11 Pro with Python~3.9.25, PyTorch~2.8.0+cu128, and CUDA~12.8. Our encoder uses seed 0 for initialization and seed 1 for stratified sampling. Each configuration was timed once with a single job resident and AC power; consequently, we do not interpret timing differences of only a few percent.

\subsection{Baselines}
\label{sec:baselines}

The grid-seeded encoder serves as the primary baseline. We rescore its stored published models where available and use one additional run of each relevant configuration to record its optimization time under its original protocol. For Miranda, we rescore the stored 200K model trained on the full-resolution grid and decode it on the stride-3 evaluation lattice. For HACC, NeurComp~\cite{lu2021compressive} and a multiresolution hash-grid representation~\cite{muller2022instant} are fitted to the $256^3$ particle deposit. We use the public W-VEG implementation and retain its released 16{,}000 iteration schedule, false-negative and false-positive regularization weights of 0.5, and inactive scale cap. To provide regular-grid inputs through the release's precomputed-samples path, we convert each grid to an explicit hexahedral \texttt{.vtu} mesh and generate precomputed samples using an external implementation of W-VEG's tetrahedral sampler. We set the primitive cap to 32{,}000 for Vortex and 200{,}000 for Miranda; only Miranda also changes the initialization fraction, to 0.00325, whereas Vortex retains the release default. W-VEG's final counts exceed the requested caps by approximately 1{,}000 primitives and define the matched comparison budgets: 33{,}000 on Vortex and 201{,}000 on Miranda. W-VEG models are decoded to the reference lattice with the authors' renderer and verified against the release training log and evaluator. To characterize relative optimization-loop cost on the same laptop-GPU platform, all timing comparisons use locally measured wall-clock times rather than W-VEG’s published A100 GPU timings. W-VEG’s public pipeline operates on structured or unstructured meshes. We therefore evaluate it on regular-grid inputs through its supported mesh-based preprocessing path; evaluating HACC would first convert the native particles to a mesh or deposited grid.

\subsection{Evaluation}
\label{sec:evaluation}

Voxel peak signal-to-noise ratio (PSNR)~\cite{wang2009mse} is the primary scalar-field metric, which summarizes the pointwise error in the normalized reconstructed scalar field; higher dB values indicate lower mean-squared error. At every voxel center of the reference lattice, the sample-based and grid-seeded models are decoded by Eq.~\ref{eq:decoded_field}, including its 0.05 mixture-density floor, and compared with the normalized reference field over $[0,1]$. Likewise, W-VEG is decoded to the reference lattice using the authors’ renderer, and the resulting grid is scored by the same evaluator. Sampled PSNR evaluates 500{,}000 random voxel centers for the sample-based and grid-seeded models with the same 0.05 floor; it is the grid-seeded encoder’s original evaluation metric. For temporal convergence, we use the PSNR logged by the sample-based trainer on its validation samples. Rendered-image PSNR and the structural similarity index measure (SSIM)~\cite{wang2004ssim} compare VTK GPU ray casts~\cite{schroeder2006vtk} of each reconstructed grid and its reference under the same saved camera, transfer function, lighting, and data range. Rendered-image PSNR summarizes pixel-wise image error, whereas SSIM measures similarity in luminance, contrast, and structure; higher PSNR and an SSIM value nearer 1 indicate greater image similarity. Bubble Plume is evaluated within the 1.38--4.27 window, Deep Water Impact scenes clip the ocean below the surface, and grid-seeded Vortex panels retain that model's saved render-time density mapping. As a decoder check, the stored 32K Vortex model scores 28.32\,dB, consistent with its published 28.3\,dB, and a repeated 200K Bubble Plume run scores 38.66\,dB, consistent with its published 38.7\,dB. All reported sizes are on-disk sizes. For our Gaussian encodings, float32 files are uncompressed archives of 48 bytes per primitive plus metadata, whereas 8-bit exports use mixed-precision quantization followed by lossless archive compression.

\section{Results}
\label{sec:results}

We evaluate the sample-based formulation by input form. HACC distinguishes matched particle-initialized controls from models supervised at a finer particle-deposit resolution. Regular-grid experiments compare the sample-based encoder with the grid-seeded encoder of Martin et al.~\cite{martin2026fixedbudget} using the same input grid, primitive budget, encoded size, and decoder. Two grids additionally compare with W-VEG~\cite{dyken2026wveg} under the protocol of Section~\ref{sec:baselines}. The remaining subsections evaluate native unstructured-cell input, temporal warm starts, and encoding cost.

\begin{table}[h]
\centering
\vspace{-2.5pt}
\caption{HACC rank-0 gas density evaluated against the exact $256^3$ particle deposit (PSNR in dB) under one camera and transfer function. The grid-seeded encoder and neural baselines are fitted to the $256^3$ deposit. All sample-based rows are initialized from the same native SPH particles. Unmarked rows use the $1024^3$ particle deposit for fidelity supervision, coverage, and relocation; rows marked $^{\ast}$ use the $256^3$ deposit fitted by the grid-seeded and neural baselines.}
\vspace{-5pt}
\label{tab:hacc}
\footnotesize
\setlength{\tabcolsep}{2pt}
\begin{tabular}{lrrrr}
\toprule
Model & Size (MB) & Voxel PSNR & Image PSNR & SSIM \\
\midrule
Grid-seeded, 300K & 14.4 & 28.06 & 28.7 & 0.895 \\
Sample-based, 300K, 8-bit & 3.4 & 32.67 & 36.5 & 0.975 \\
Sample-based, 300K, 8-bit$^{\ast}$ & 3.5 & 44.72 & 42.5 & 0.992 \\
Sample-based, 650K & 31.2 & 35.24 & 36.6 & 0.976 \\
Sample-based, 650K, 8-bit & 7.0 & 35.22 & 36.6 & 0.976 \\
Sample-based, 650K$^{\ast}$ & 31.2 & 44.62 & 41.6 & 0.990 \\
NeurComp~\cite{lu2021compressive} & 1.1 & 39.43 & 35.2 & 0.954 \\
Hash grid~\cite{muller2022instant} & 43.5 & 62.61 & 60.5 & 0.9998 \\
\bottomrule
\end{tabular}
\vspace{-5pt} 
\end{table}

\subsection{Particle Data: HACC}
\label{sec:res_hacc}

\begin{figure}[b]
\vspace{-10pt} 
\centering
\includegraphics[width=\linewidth,alt={Scatter plot of HACC rendered PSNR in decibels versus model size on disk in megabytes on a logarithmic horizontal axis. Blue markers are neural representations, orange is the grid-seeded Gaussian encoder, and green markers are sample-based Gaussian encoders; circles and squares indicate 1024-cubed and 256-cubed supervision, and hollow markers are 8-bit exports. An arrow compares equal-budget, 256-cubed-supervision models: the 3.5 MB sample-based export reaches 42.5 dB, versus 28.7 dB for the 14.4 MB grid-seeded model. The 43.5 MB hash grid reaches 60.5 dB.}
]{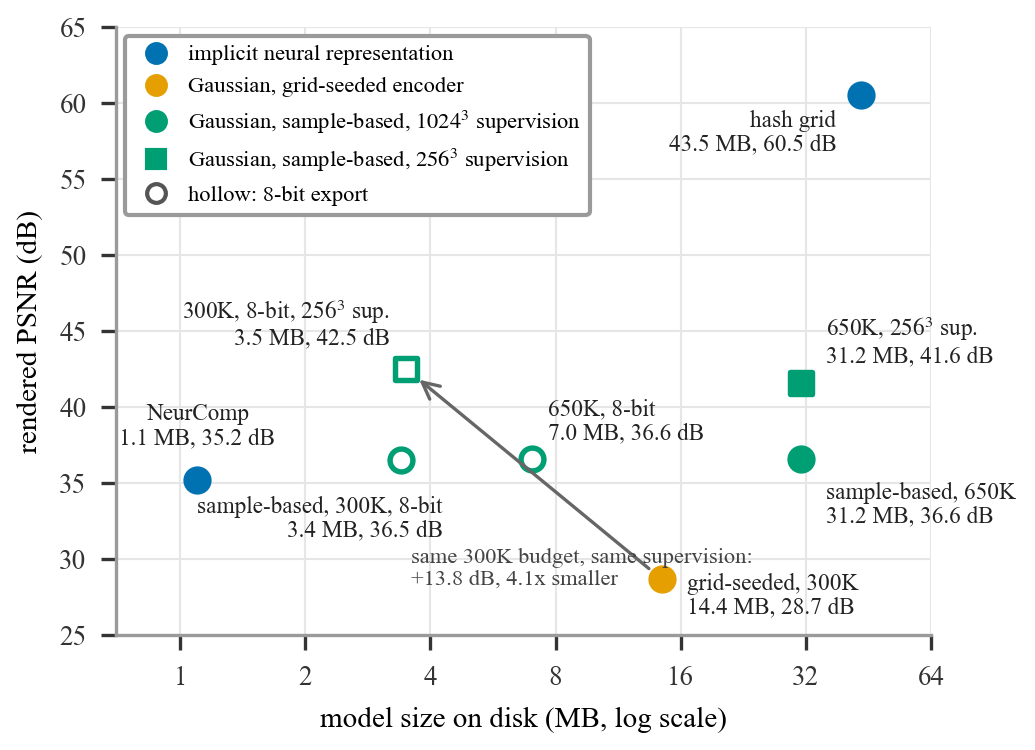}
\vspace{-16pt} 
\caption{HACC rank-0 gas-density size--fidelity trade-off; Rendered PSNR plotted against model size on disk (log scale); Colors identify implicit neural (blue), grid-seeded Gaussian (orange), and sample-based Gaussian (green) representations; green circles and squares denote $1024^3$ and $256^3$ supervision, respectively, and hollow markers denote 8-bit exports. Arrow highlights the matched 300K, $256^3$-supervision comparison: sample-based export improves rendered PSNR by 13.8~dB while reducing stored size by $4.1\times$.}
\label{fig:hacc_chart}
\end{figure}

\begin{figure}[b]
\vspace{-10pt} 
\centering
\includegraphics[width=\linewidth,alt={Two rows of HACC particle-property renderings. The top row compares the 256-cubed gas-temperature deposit with a 300K float32 sample-based encoding and its 8-bit export; the bottom row makes the same comparison for metal mass fraction. The 8-bit reconstructions visually preserve the diffuse temperature structures and compact metal-enriched regions, with rendered PSNR changing from 52.5 to 51.7 dB for temperature and from 38.9 to 38.8 dB for metal mass fraction.}
]{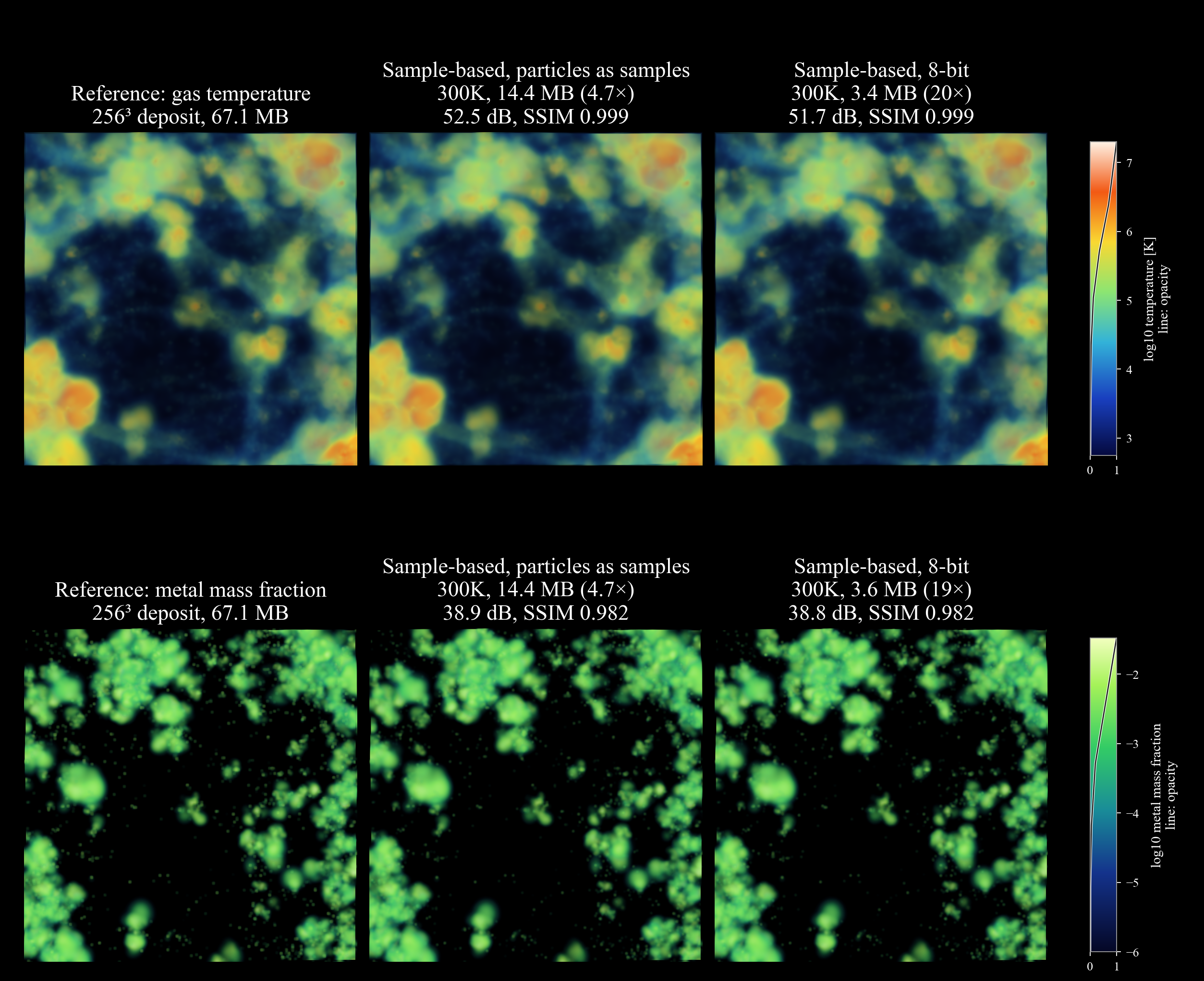}
\vspace{-12.5pt} 
\caption{Two further HACC particle properties encoded from the same particles with the sample-based encoder at 300K (4{,}000 iterations, $256^3$ supervision): gas temperature (top) and metal mass fraction (bottom), each as the $256^3$ deposit, the float32 model, and its 8-bit export, labeled with size, reduction factor, and rendered PSNR/SSIM under the field's own transfer function.}
\label{fig:hacc_props}
\vspace{-10pt} 
\end{figure}

Table~\ref{tab:hacc} evaluates HACC rank-0 gas density against the exact $256^3$ particle deposit. All sample-based HACC models are initialized from the same native SPH particles; the starred rows use the $256^3$ deposit for fidelity supervision, coverage, and relocation, matching the supervision grid used by the grid-seeded encoder and both neural baselines. Under this controlled $256^3$-supervision setting, the 300K 8-bit sample-based encoding reaches 44.72\,dB voxel PSNR and 42.5\,dB image PSNR, compared with 28.06\,dB and 28.7\,dB for the 300K grid-seeded encoding. Its 3.5\,MB 8-bit export is $4.1\times$ smaller than the 14.4\,MB float32 grid-seeded model. The unmarked sample-based rows use the finer $1024^3$ particle deposit for field-space supervision and are therefore reported as a separate supervision-resolution configuration rather than as controlled replacements for the $256^3$-trained baselines. For the 8-bit exports in this configuration, increasing the budget from 300K to 650K increases voxel PSNR from 32.67 to 35.22\,dB, while image PSNR changes from 36.5 to 36.6\,dB. The 650K 8-bit export reduces stored size from 31.2\,MB for the float32 model to 7.0\,MB, with a 0.02\,dB voxel-PSNR change. Figure~\ref{fig:hacc} visualizes the HACC reconstructions (top panel) and their absolute luminance differences (bottom panel) under the same camera and transfer function. Both particle-initialized sample-based models retain the small halos as separate features, whereas the grid-seeded reconstruction broadens them into neighboring filaments and produces the largest spatially distributed residual among the Gaussian encodings. The neural baselines provide size--fidelity reference points: NeurComp preserves the overall filamentary pattern but retains visible error around bright structures, while the hash grid matches the reference most closely but at a substantially larger on-disk size of 43.5~MB. The $1024^3$-supervision sample-based export attains higher rendered PSNR than NeurComp (36.5 versus 35.2~dB) and a lower mean absolute luminance difference (0.0077 versus 0.0084), while the controlled $256^3$-supervision sample-based export further lowers the mean absolute luminance difference to 0.0035 and appears visually close to the hash-grid reconstruction in this view while demonstrating significant saved storage of 3.5~MB compared to Hash grid's 43.5~MB. Figure~\ref{fig:hacc_chart} summarizes the HACC size--fidelity trade-off using rendered PSNR and on-disk model size. The annotated pair isolates the controlled comparison: at the same 300K primitive budget and $256^3$ supervision, the 3.5~MB sample-based export reaches 42.5~dB, compared with 28.7~dB for the 14.4~MB grid-seeded model. Thus, the sample-based encoding provides 13.8~dB higher rendered PSNR while using 4.1$\times$ less storage. NeurComp provides a smaller 1.1~MB neural reference point but is 7.3~dB below the controlled sample-based export, whereas the hash grid reaches 60.5~dB at 43.5~MB, more than twelve times the controlled export's stored size. These neural results therefore contextualize the size--fidelity trade-off rather than constitute primitive-budget-matched comparisons. Figure~\ref{fig:hacc_props} demonstrates the extension of particle-initialized representation beyond gas density to gas temperature (top panel) and metal mass fraction (bottom panel). For temperature, the 14.4~MB float32 model reaches 52.5~dB rendered PSNR (47.4~dB voxel PSNR), and the 3.4~MB 8-bit export retains the diffuse temperature structures with no visually material loss at the displayed scale, decreasing rendered PSNR by 0.8~dB to 51.7~dB; both results report SSIM 0.999, indicating very high rendered structural similarity under this visualization setting. For the sparser metal mass-fraction field, the 8-bit export preserves the compact enriched regions while reducing stored size from 14.4~MB to 3.6~MB; rendered PSNR changes only from 38.9 to 38.8~dB, and both models report the same SSIM 0.982. Hence, the 8-bit exports provide 20$\times$ reduction for temperature and 19$\times$ reduction for metal mass fraction relative to their 67.1~MB deposits, while significantly retaining the visual structures and rendered fidelity of the float32 encodings.

\subsection{Regular Grids as Samples}
\label{sec:res_grids}

\begin{figure}[h]
\centering
\includegraphics[width=0.49\textwidth,
  alt={Two regular-grid comparisons, each with reference, grid-seeded encoder, and sample-based encoder at equal primitive budget and file size. Top: Vortex at 32K Gaussians; the sample-based result preserves separated tubular structures that are broadened by the grid-seeded result. Bottom: Bubble Plume at 200K Gaussians; the sample-based result preserves separation between the outer sheath and internal plume filaments.}
]{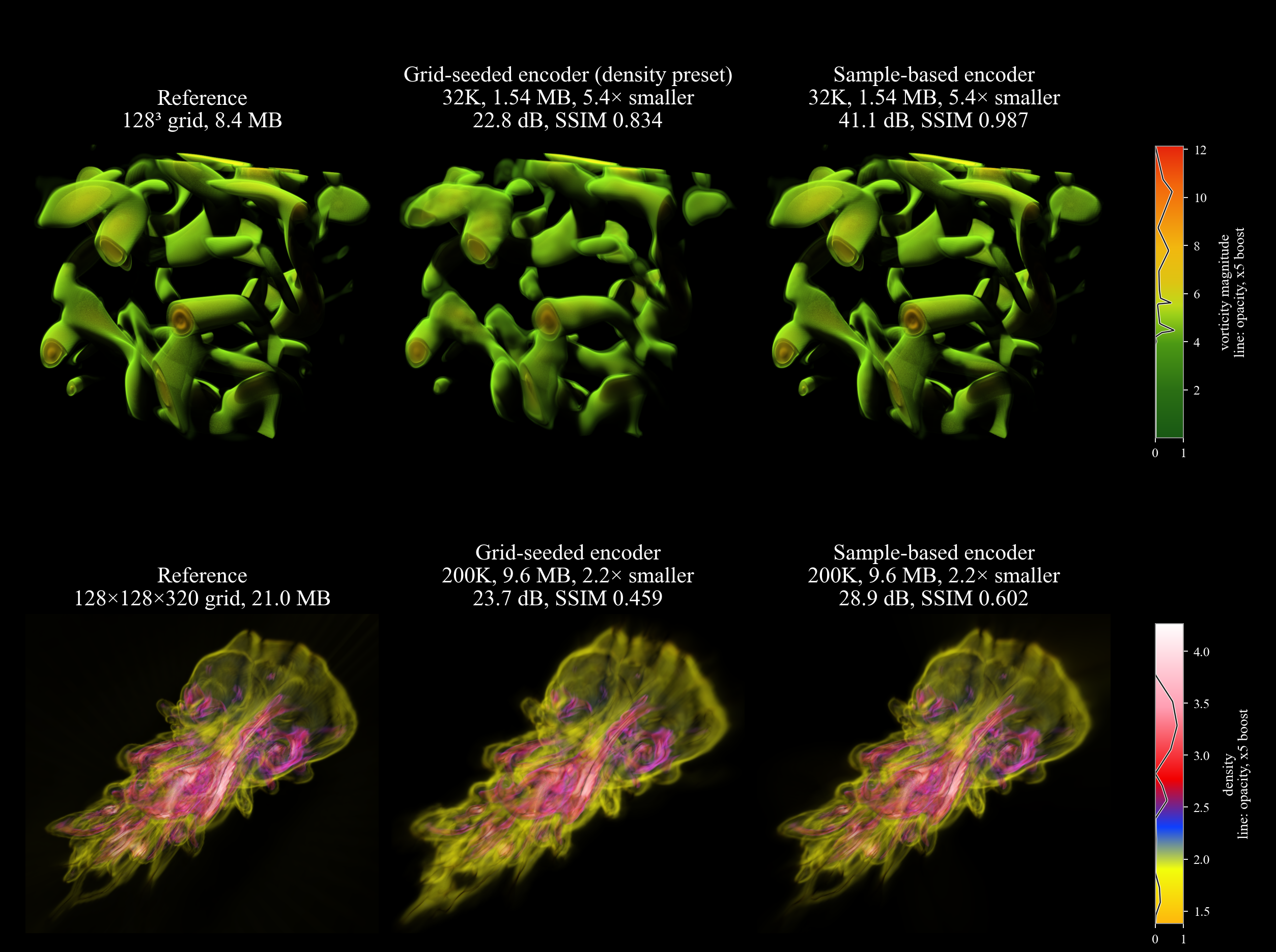}
\vspace{-12.5pt} 
\caption{Regular grids as samples at identical budget and file size, ray cast from the decoded grids under each dataset's saved transfer function, camera, and lighting (bar: color over value, line: opacity). Top: Vortex at 32{,}000 Gaussians (1.54\,MB); the grid-seeded panel uses the density mapping saved for that model, and the rendered PSNR rises by 18.3\,dB. Bottom: Bubble Plume (stride-2 grid, $128{\times}128{\times}320$) at 200{,}000 Gaussians (9.6\,MB) under the saved data window; the rendered PSNR rises by 5.2\,dB, and the filaments inside the plume stay separate.}
\label{fig:grids}
\vspace{-10pt}
\end{figure}

Table~\ref{tab:grids} compares the two encoders when both receive the same grid voxels and use the same primitive budget, stored size, and decoder. The sample-based encoder improves voxel PSNR by 21.35\,dB on Vortex at 32K primitives and by 6.18\,dB on Bubble Plume at 200K primitives within the saved 1.38--4.27 data window. For scale context beyond the matched-budget comparison in Table~\ref{tab:grids}, the 32K sample-based model reaches 49.67\,dB, exceeding the 44.90\,dB reported for the largest published grid-seeded model, which uses 1.4M primitives~\cite{martin2026fixedbudget}; this comparison corresponds to approximately $44\times$ fewer primitives. Figure~\ref{fig:grids} shows the corresponding renderings under the saved visualization settings. On Vortex, rendered PSNR increases from 22.8 to 41.1\,dB and SSIM from 0.834 to 0.987. The sample-based reconstruction preserves the separated tubes visible in the reference, whereas the grid-seeded reconstruction broadens their cross sections. On Bubble Plume, rendered PSNR increases from 23.7 to 28.9\,dB and SSIM from 0.459 to 0.602. The sample-based reconstruction preserves separation between the outer sheath and inner plume filaments that merge in the grid-seeded reconstruction.

\begin{table}[h]
\centering
\vspace{-2.5pt}
\caption{Regular grids: the grid-seeded encoder and the sample-based encoder supplied with the same grid voxels, at identical primitive budget and file size. Both are decoded with Eq.~\ref{eq:decoded_field}; values are voxel PSNR in dB. Bubble Plume is evaluated within its saved data window, 1.38--4.27.}
\label{tab:grids}
\vspace{-5pt} 
\footnotesize
\setlength{\tabcolsep}{3pt}
\begin{tabular}{llrrr}
\toprule
Dataset & Budget & MB & Grid-seeded & Sample-based \\
\midrule
Vortex ($128^3$) & 32K & 1.54 & 28.32 & 49.67 \\
Bubble Plume (stride 2) & 200K & 9.6 & 40.97 & 47.15 \\
\bottomrule
\end{tabular}
\vspace{-10pt} 
\end{table}

\subsection{Evaluation at Matched Primitive Counts}
\label{sec:res_wveg}

\begin{figure*}[ht]
\centering
\includegraphics[width=0.94\textwidth,
  alt={Two-row Vortex comparison under one transfer function, camera, and lighting. Top row: reference, W-VEG at 33K primitives, full-schedule and matched-schedule 33K sample-based models, and a 32K grid-seeded model. The full-schedule sample-based model most closely retains the smooth tube boundaries and internal voids; W-VEG has fragmented tube surfaces, and the grid-seeded model broadens the tubes. Bottom row: the 32K grid-seeded model at 1500 iterations and sample-based models at 1500, 500, and 300 iterations, all with the same 1.54 MB stored size; the sample-based models retain the tube structure at shorter optimization times.}
]{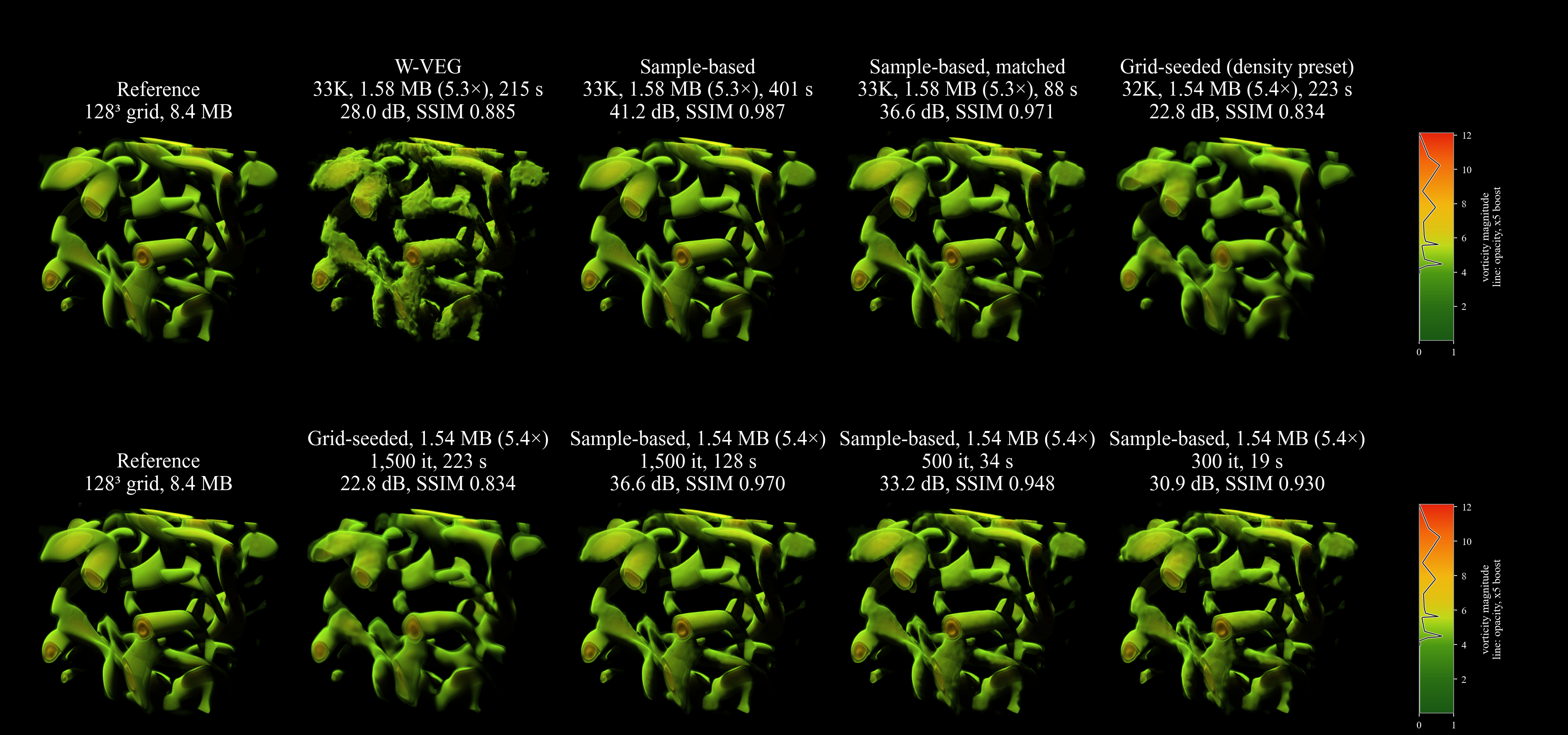}
\vspace{-2.5pt} 
\caption{Vortex reconstructions under the saved transfer function, camera, and lighting; grid-seeded panels use that model's saved density mapping. Top: the reference, W-VEG's 33K release model, full- and matched-schedule 33K sample-based models, and the 32K grid-seeded model. Bottom: the 32K grid-seeded model and sample-based models at the matched 1,500-iteration schedule, 500-, and 300-iteration short schedules. Panel labels report stored size, schedule, optimization time, rendered PSNR, and SSIM.}
\label{fig:wveg_vortex}
\vspace{-10pt} 
\end{figure*}

\begin{figure*}[tbp]
\centering
\includegraphics[width=\textwidth,
  alt={Stride-3 Miranda scalar-field renderings at approximately 200K primitives: reference, W-VEG at 201K primitives, full-schedule sample-based encoding, matched-schedule sample-based encoding, and the grid-seeded encoder's stored 200K model. The full-schedule sample-based result most closely matches the large turbulent structures and localized bright features of the reference. The grid-seeded result shows visible speckling along the left side of the volume.}
]{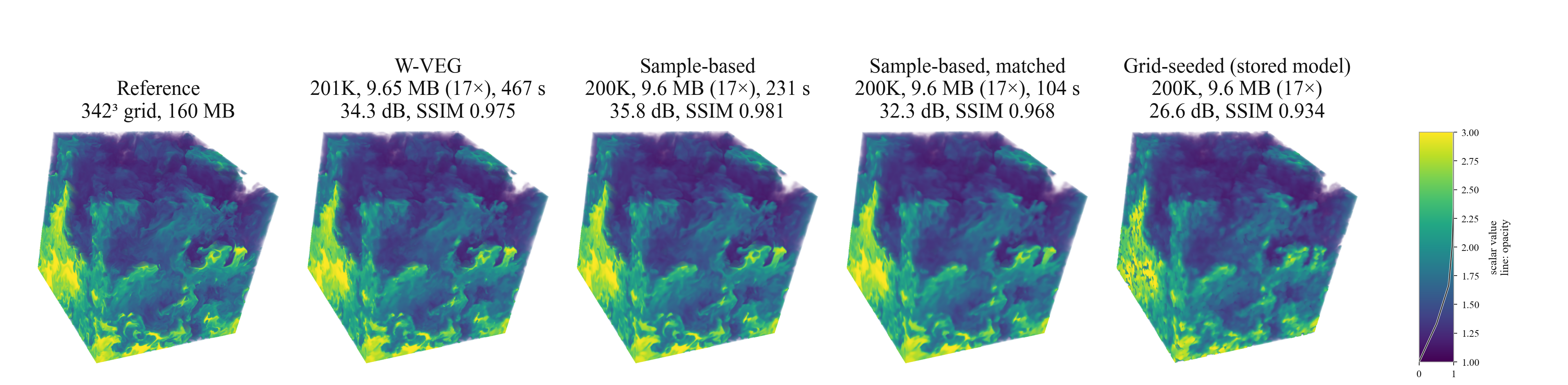}
\vspace{-20pt} 
\caption{Stride-3 Miranda ($342^3$) at $\sim$200K primitives under the protocol of Figure~\ref{fig:wveg_vortex} (top); the grid-seeded panel is the grid-seeded encoder's stored model, trained on the full grid and decoded on the stride-3 lattice. On this dense field the sample-based encoder renders 1.42\,dB above W-VEG.}
\label{fig:wveg_miranda}
\vspace{-10pt} 
\end{figure*}

Table~\ref{tab:wveg} evaluates the sample-based encoder at matched or near-matched primitive counts against W-VEG’s final release-schedule model and also reports the grid-seeded encoder. Each reconstructed field is evaluated on the same reference lattice after method-specific decoding, as described in Section~\ref{sec:baselines}. On Vortex, the full-schedule sample-based encoder reaches 49.72\,dB voxel PSNR, 12.45\,dB above W-VEG at 33K primitives. The matched 1{,}500-iteration sample-based schedule reaches 45.14\,dB, remaining 7.87\,dB above W-VEG while using 41\% of its optimization time. On stride-3 Miranda, the full-schedule sample-based encoder reaches 29.74\,dB, 2.28\,dB above W-VEG at approximately 200K primitives while using 231\,s rather than 467\,s. The matched schedule reaches 26.72\,dB, 0.74\,dB below W-VEG while using 22\% of its optimization time. The top row in figure~\ref{fig:wveg_vortex} visualizes the Vortex comparisons at nearly equal on-disk size: The full-schedule sample-based model preserves the smooth tubular boundaries and internal voids of the reference, while the matched schedule retains the principal tubes at substantially lower optimization time (88 s rather than 401 s). With this memory constraint, W-VEG exhibits irregular and fragmented tube surfaces, whereas the grid-seeded reconstruction broadens the tubes under its saved density mapping. The matched sample-based model reaches 36.6~dB rendered PSNR, 8.6~dB above W-VEG's 28.0~dB under this view. Figure~\ref{fig:wveg_miranda} shows the Miranda reconstructions, where in both cases, the full-schedule sample-based reconstruction more closely retains the reference's visible structures than the matched schedule, while the grid-seeded Miranda model exhibits a distinct speckled texture resolved by the sample-based encoder.

\begin{table}[h]
\centering
\vspace{-2.5pt}
\caption{Comparison with W-VEG~\cite{dyken2026wveg} at its final primitive count (PSNR in dB). W-VEG uses the public implementation under its 16{,}000-iteration release schedule and is decoded onto the reference lattice with the authors' renderer. The sample-based and grid-seeded encoders use Eq.~\ref{eq:decoded_field}. Every reconstructed lattice is scored by the same evaluator; times are wall-clock optimization times on the RTX~5090 laptop GPU. The grid-seeded Vortex row is the published 32K model, whereas the grid-seeded Miranda row is the grid-seeded encoder's stored model, trained on the full-resolution grid and decoded on the stride-3 lattice.}
\label{tab:wveg}
\vspace{-5pt} 
\footnotesize
\setlength{\tabcolsep}{3pt}
\begin{tabular}{lrrrr}
\toprule
Method & Primitives & Time (s) & Voxel & Image / SSIM \\
\midrule
\multicolumn{5}{l}{\emph{Vortex, $128^3$, budget 33K}} \\
W-VEG, release schedule & 33{,}000 & 215 & 37.27 & 28.0 / 0.885 \\
Sample-based, own & 33{,}000 & 401 & 49.72 & 41.2 / 0.987 \\
Sample-based, matched & 33{,}000 & 88 & 45.14 & 36.6 / 0.971 \\
Grid-seeded & 32{,}000 & 223 & 28.32 & 22.8 / 0.834 \\
\addlinespace
\multicolumn{5}{l}{\emph{Miranda, stride 3 ($342^3$), budget 200K}} \\
W-VEG, release schedule & 201{,}000 & 467 & 27.46 & 34.35 / 0.975 \\
Sample-based, own & 200{,}000 & 231 & 29.74 & 35.77 / 0.981 \\
Sample-based, matched & 200{,}000 & 104 & 26.72 & 32.30 / 0.968 \\
Grid-seeded, stored & 199{,}979 & --- & 20.92 & 26.60 / 0.934 \\
\bottomrule
\end{tabular}
\vspace{-10pt} 
\end{table}

\subsection{Unstructured Cells: Deep Water Impact}
\label{sec:res_unstructured}

\begin{figure*}[ht]
\centering
\includegraphics[width=1\textwidth,
  alt={Deep Water Impact renderings. Top row compares the volume-weighted deposit of 44.9 million adaptive hexahedral cells at cycle 12,244 with a 400K sample-based encoding initialized from the native cells; insets magnify the splash. Both show a thin sea sheet and localized splash. Bottom row compares a three-field composite at cycle 17,260 for the reference, independently encoded fields, and warm-started fields with relocation. Water fraction is cyan, temperature is yellow, and asteroid fraction is pink.}
]{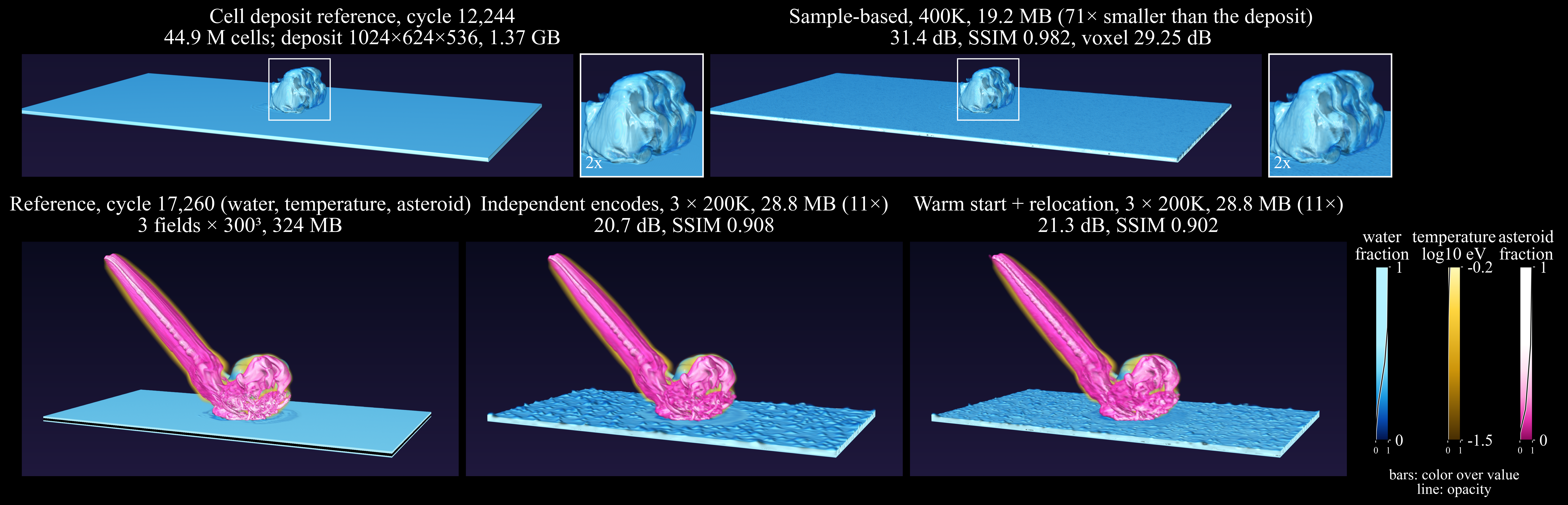}
\vspace{-12.5pt} 
\caption{Deep Water Impact, ray cast with the series' transfer functions and cropped to the sea surface. Top: cycle 12{,}244, volume-weighted deposit of the 44.9\,M cells and 400K-Gaussian encoding seeded directly from native cells and refined against this deposit, with the splash magnified in the inset. Bottom: cycle 17{,}260 of the $300^3$ series with the water fraction, temperature, and asteroid fraction each encoded at 200K Gaussians (28.8\,MB total) and composited: the reference, independent encodes, and warm starts with relocation (bars: color over value, line: opacity).}
\label{fig:dwi}
\vspace{-10pt} 
\end{figure*}

We encode cycle 12{,}244 of Deep Water Impact from 44{,}851{,}632 adaptive-mesh hexahedral cells using 400K Gaussians. The primitives are initialized directly from the native cells and refined against the volume-weighted cell deposit defined in Eq.~\ref{eq:cell_deposit}. Thus, the cells determine primitive placement and initial scale, while the materialized deposit provides the common field-space supervision and evaluation target; no resampled grid is used to seed the primitive set. After 3{,}000 iterations, the 19.2\,MB model reaches 29.25\,dB voxel PSNR against the deposited reference and 31.4\,dB rendered PSNR with SSIM 0.982 under the shaded-water transfer function. The top row of Figure~\ref{fig:dwi} shows that the reconstruction preserves both the sea sheet and the localized splash, with the visible residual concentrated near the sea-surface seam where coarse and fine cells meet.

\begin{table}[h]
\centering
\caption{Water fraction of the Deep Water Impact series on the $300^3$ grid, 200K Gaussians: sampled PSNR (dB) on the trainer's validation points. Independent encodes were seeded from the step's own voxels and ran 3{,}000 iterations (262--317\,s); warm starts were initialized from the previous stored step's independent model and ran 1{,}000 iterations without relocation on the coarse stride and 1{,}500 iterations with relocation on the dense stride (143--163\,s).}
\vspace{-5pt} 
\label{tab:temporal}
\footnotesize
\begin{tabular}{rrrr}
\toprule
Cycle & Stride (cycles) & \shortstack{Independent final\\PSNR (dB)} & \shortstack{Warm final\\PSNR (dB)} \\
\midrule
\multicolumn{4}{l}{\emph{Coarse stride, no relocation, 1{,}000 warm iterations}} \\
11{,}642 & 1{,}860 & 28.93 & 25.18 \\
13{,}306 & 1{,}664 & 28.55 & 26.32 \\
14{,}509 & 1{,}203 & 28.50 & 26.65 \\
15{,}422 & 913 & 28.93 & 27.11 \\
16{,}118 & 696 & 29.62 & 27.93 \\
\addlinespace
\multicolumn{4}{l}{\emph{Dense stride, relocation on, 1{,}500 warm iterations}} \\
16{,}709 & 591 & 29.84 & 29.99 \\
17{,}260 & 551 & 30.03 & 30.03 \\
17{,}777 & 517 & 29.77 & 29.78 \\
18{,}278 & 501 & 30.04 & 29.87 \\
\bottomrule
\end{tabular}
\vspace{-10pt} 
\end{table}

\subsection{Temporal Reuse and Primitive Relocation}
\label{sec:temporal}

\begin{figure}[tbp]
\centering
\includegraphics[width=\linewidth,
  alt={Plot of validation-sample PSNR versus stride from the previous stored Deep Water Impact cycle, with the encoded cycle shown beneath each tick. Filled black circles are independent 3000-iteration encodes. Orange hollow circles are 1000-iteration warm starts without relocation at coarse strides and finish below the independent results. Green hollow circles are 1500-iteration warm starts with relocation at dense strides and finish near the independent results; green triangles mark their iteration-1100 peaks.}
]{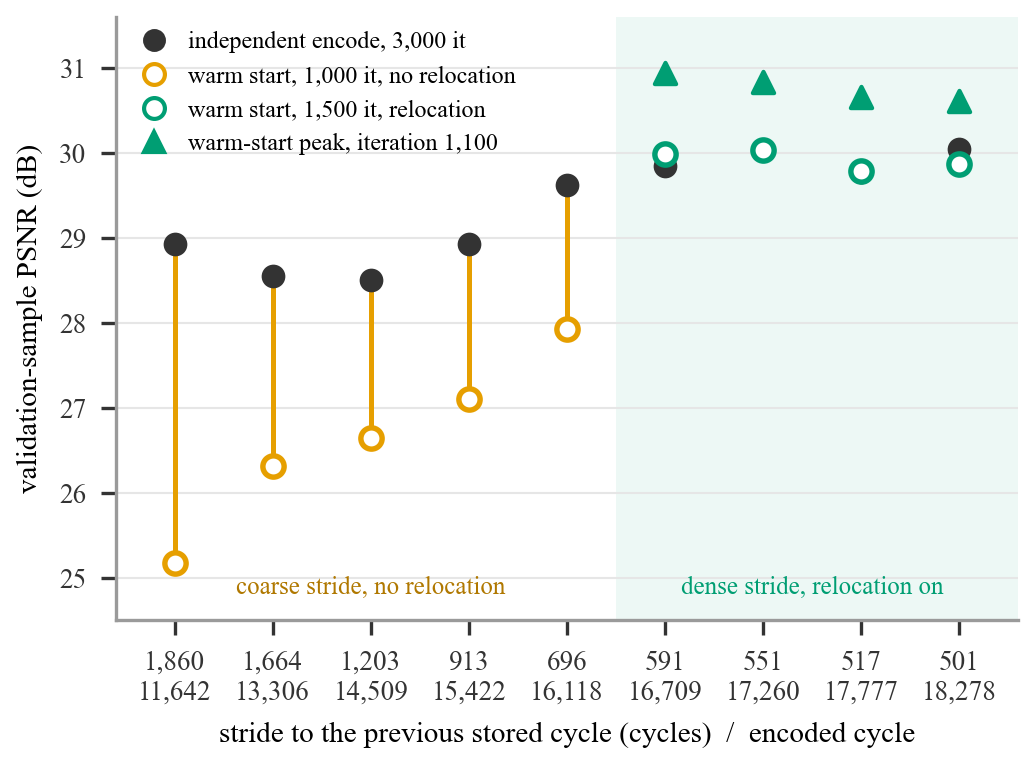}
\vspace{-15pt} 
\caption{Warm starts vs. independently encoded water-fraction models at 200K Gaussians: validation-sample PSNR vs. stride to the previous stored cycle, with the encoded cycle beneath each tick. Filled markers show the independent final after 3{,}000 iterations; hollow markers show the warm-start final after 1{,}000 iterations without relocation on coarse strides or 1{,}500 iterations with relocation on dense strides; triangles show dense-stride warm-start peak at iteration 1{,}100. Dense-stride warm starts reach/exceed the independent final during refinement and finish within 0.17\,dB.}
\vspace{-15pt}
\label{fig:temporal}
\end{figure}

The Deep Water Impact series contains 26 stored cycles; we evaluate water-fraction encodings at nine cycles on the $300^3$ grid, using 200K Gaussians per step. Each independent model is seeded from that step's voxels and optimized for 3{,}000 iterations. Each warm start is initialized from the preceding step's independently encoded model and then optimized under the stride-dependent schedule of Table~\ref{tab:temporal}. The table reports the trainer's fixed validation-sample PSNR, rather than the full-grid voxel PSNR reported for the non-temporal reconstruction comparisons. For coarse strides of 696--1{,}860 cycles, the 1{,}000-iteration warm starts run without relocation and remain below their corresponding independent final. The final gap decreases from 3.75\,dB at the largest stride to 1.69\,dB at the smallest. For dense strides of 501--591 cycles, the 1{,}500-iteration warm starts retain relocation. Each reaches its independently trained final near iteration 800, peaks 0.6--1.1\,dB above that final near iteration 1{,}100, and finishes within 0.17\,dB of it. Figure~\ref{fig:temporal} visualizes this series-specific transition. At coarse strides of 696--1{,}860 cycles, warm starts without relocation finish 1.69--3.75\,dB below the independent encodes, shown by the orange vertical gaps. At dense strides of 501--591 cycles, relocation-enabled warm starts reach or exceed the independent final during refinement and finish within 0.17\,dB; triangles show their iteration-1{,}100 peaks. Figure~\ref{fig:dwi}'s bottom row shows cycle 17{,}260 with water fraction, temperature, and asteroid fraction encoded separately at 200K Gaussians each and composited under the series transfer functions. For water fraction, the warm start increases voxel PSNR from 31.6 to 34.0\,dB. At the same cycle, warm starts for temperature and asteroid fraction remain below their independently encoded counterparts, at 51.5 versus 55.1\,dB and 41.1 versus 44.8\,dB, respectively. For the composite, rendered PSNR increases from 20.7\,dB for the independent encodes to 21.3\,dB for the warm starts, whereas SSIM decreases from 0.908 to 0.902. These field-dependent outcomes show that temporal reuse depends on both the stride and the evolution of the encoded scalar field.

\begin{table}[h]
\centering
\caption{Data reduction achieved on every dataset: reference grid and its size on disk, the model, its size, the reduction factor, and that model's quality (voxel PSNR in dB against the reference grid; rendered PSNR in dB under the saved view). Float32 models unless marked 8-bit; the 8-bit export reduces a float32 model a further $4\times$ (measured on HACC, Table~\ref{tab:hacc}). $^{c}$~$256^3$ supervision; $^{w}$~windowed voxel PSNR (1.38--4.27); $^{s}$~sampled PSNR.}
\vspace{-5pt} 
\label{tab:reduction}
\scriptsize
\setlength{\tabcolsep}{2pt}
\begin{tabular}{@{}llrrrr@{}}
\toprule
Dataset, reference & Model & MB & Ratio & Voxel & Image \\
\midrule
HACC, $256^3$ (67.1 MB) & 300K, 8-bit$^{c}$ & 3.5 & 19$\times$ & 44.7 & 42.5 \\
Vortex, $128^3$ (8.4 MB) & 32K & 1.54 & 5.4$\times$ & 49.7 & 41.1 \\
 & 10K & 0.49 & 17$\times$ & 42.6 & 34.7 \\
Bubble Plume, stride 2 (21.0 MB) & 200K & 9.6 & 2.2$\times$ & 47.2$^{w}$ & 28.9 \\
 & 50K & 2.4 & 8.7$\times$ & 44.4$^{w}$ & 25.6 \\
Miranda, $342^3$ (160 MB) & 200K & 9.6 & 17$\times$ & 29.7 & 35.8 \\
DWI cells, deposit (1.37 GB) & 400K & 19.2 & 71$\times$ & 29.3 & 31.4 \\
DWI step, $300^3$ (108 MB) & 200K & 9.6 & 11$\times$ & 30.0$^{s}$ & 20.8 \\
DWI three fields (324 MB) & $3{\times}$200K & 28.8 & 11$\times$ & --- & 20.7 \\
\bottomrule
\end{tabular}
\vspace{-7.5pt} 
\end{table}

\subsection{Encoding Cost and Short Schedules}
\label{sec:res_cost}

Table~\ref{tab:reduction} collects the reduction factor reached on every dataset together with the quality of the model that reaches it. Table~\ref{tab:cost} reports single-run optimization-loop times on the RTX~5090 laptop GPU; the timing scope and its limitations are specified in Section~\ref{sec:training}. For the 32K Vortex and 200K Bubble Plume comparisons in Table~\ref{tab:cost}, the matched training-budget protocol uses 1{,}500 iterations and 3{,}072 points per iteration. The sample-based encoder records 128\,s on Vortex and 91\,s on Bubble Plume, compared with 223\,s and 387\,s for the grid-seeded encoder. These correspond to approximately $1.7\times$ and $4.3\times$ lower optimization-loop times. The sample-based model achieves higher full-grid voxel PSNR on Vortex and higher sampled PSNR on Bubble Plume. On Bubble Plume, rendered PSNR is also higher, 24.1 versus 23.7\,dB. Because the two encoders retain different objectives and initialization procedures, this protocol matches iteration and sampling budgets rather than all optimization details. The HACC rows use substantially different schedules, 300 iterations for the grid-seeded encoder and 4{,}000 for the sample-based encoder. They are therefore reported as cost measurements rather than a time-to-quality comparison. Short schedules reveal the quality-time trade-off for the sample-based formulation. The 19\,s and 34\,s Vortex schedules, and the 55\,s Bubble Plume schedule, each exceed the corresponding 1{,}500-iteration grid-seeded result in full-grid voxel PSNR for Vortex or sampled PSNR for Bubble Plume. The 16\,s Bubble Plume schedule falls below the grid-seeded result in sampled PSNR, windowed voxel PSNR, and rendered PSNR. The two sub-minute Bubble Plume schedules trail the grid-seeded rendering by 0.9 and 2.5\,dB, respectively. Reducing the primitive budget provides an additional operating point. At 10K primitives, Vortex reaches 42.59\,dB voxel PSNR with a 0.49\,MB model, exceeding the 32K grid-seeded model by 14.27\,dB at approximately one-third of its file size. At 50K primitives, Bubble Plume reaches 40.27\,dB sampled PSNR and 44.41\,dB windowed voxel PSNR, exceeding the 200K grid-seeded model by 1.61 and 3.44\,dB, respectively, at one-quarter of the file size. Figure~\ref{fig:wveg_vortex}'s bottom row shows that the 19\,s Vortex encoding remains 8.1\,dB above the 223\,s grid-seeded rendering while retaining the reference tubes. For reference, constructing samples from the $128^3$ Vortex grid takes 0.22\,s and seeding 32{,}000 primitives takes less than 0.01\,s. 

\section{Discussion}
\label{sec:discussion}

\subsection{Representation Behavior}
\label{sec:discussion_behavior}

The regular-grid comparisons evaluate the practical effect of replacing grid-derived allocation with sample-based initialization and refinement at the same primitive budget, stored size, and decoder. The resulting gains are consistent with the sample-based objective concentrating Gaussian support on localized field structure while retaining only a coverage constraint on mixture density. Unlike the grid-seeded encoder~\cite{martin2026fixedbudget}, our sample-based objective omits density-uniformity and inter-primitive smoothness terms. This permits narrow, nonuniform support where it improves field reconstruction, while the coverage loss promotes sufficient mixture support for normalized field evaluation within occupied regions. The experiments do not separately ablate every changed component; consequently, the results establish the behavior of the full sample-based formulation rather than assigning the improvement to one loss term alone. The HACC experiments also show that the supervision reference is part of the experimental configuration. All sample-based HACC models begin from the same native particles, but Table~\ref{tab:hacc} distinguishes controls supervised on the $256^3$ deposit from models supervised on the $1024^3$ deposit. Comparisons across supervision lattices characterize distinct representation settings, as target resolution changes the optimized field, the index-space particle widths, and the structures retained by the reference. We therefore use the $256^3$-supervision rows for controlled comparisons with the grid-seeded and neural baselines, and report the $1024^3$-supervision rows as a separate particle-deposit configuration.

\begin{table}[h]
\centering
\vspace{2.5pt}
\caption{Encoding cost on the RTX~5090 laptop GPU. The matched protocol uses the same budget, 1{,}500 iterations, and 3{,}072 points per iteration for both encoders; the HACC rows use each encoder's own schedule; the short schedules use 8{,}192 points per iteration. Quality is the sampled PSNR (500{,}000 random voxels) unless marked.}
\vspace{-5pt}
\label{tab:cost}
\footnotesize
\setlength{\tabcolsep}{2pt}
\begin{tabular}{llrrr}
\toprule
Dataset, budget & Encoder & Iter. & Time (s) & PSNR (dB) \\
\midrule
\multicolumn{5}{l}{\emph{Matched protocol}} \\
Vortex, 32K & Grid-seeded & 1{,}500 & 223 & 28.3 \\
 & Sample-based & 1{,}500 & 128 & 45.1 \\
Bubble Plume, 200K & Grid-seeded & 1{,}500 & 387 & 38.7 \\
 & Sample-based & 1{,}500 & 91 & 40.0 \\
\addlinespace
\multicolumn{5}{l}{\emph{Own schedules}} \\
HACC, 300K & Grid-seeded & 300 & 65.5 & 28.06$^{\dagger}$ \\
 & Sample-based & 4{,}000 & 384 & 32.67$^{\dagger}$ \\
\addlinespace
\multicolumn{5}{l}{\emph{Short schedules}} \\
Vortex, 32K & Sample-based & 500 & 34 & 41.49$^{\dagger}$ \\
 & Sample-based & 300 & 19 & 39.11$^{\dagger}$ \\
Bubble Plume, 200K & Sample-based & 800 & 55 & 39.60 (42.80)$^{\ddagger}$ \\
 & Sample-based & 300 & 16 & 38.09 (40.07)$^{\ddagger}$ \\
\addlinespace
\multicolumn{5}{l}{\emph{Smaller budgets, matched schedule}} \\
Vortex, 10K (0.49\,MB) & Sample-based & 1{,}500 & 132 & 42.59$^{\dagger}$ \\
Bubble Plume, 50K (2.4\,MB) & Sample-based & 1{,}500 & 98 & 40.27 (44.41)$^{\ddagger}$ \\
\bottomrule
\end{tabular}
\par\vspace{2pt}
{\footnotesize $^{\dagger}$Voxel PSNR on the full grid. $^{\ddagger}$Sampled PSNR, with the windowed voxel PSNR in parentheses.\par}
\vspace{-12.5pt} 
\end{table}

\subsection{Limitations and Future Work}
\label{sec:limitations}

While particle and cell samples determine primitive initialization, the current refinement procedure materializes a reference grid: a particle deposit for HACC and a volume-weighted cell deposit for Deep Water Impact. The method therefore uses samples at the representation-input stage and a materialized reference grid during optimization. Likewise, the unstructured path currently assumes axis-aligned hexahedral cells. Future work will investigate direct particle- and cell-query supervision when a dense deposited target exceeds available memory, as well as shape-aware initialization and evaluation for tetrahedra, wedges, and general polyhedral cells.

The current implementation provides an offline, file-based workflow that loads and converts stored data products and constructs reference deposits used for particle and cell refinement. The reported times characterize the optimization loop; evaluating deployment during a simulation run will require end-to-end measurements of the transfer of fields from simulation memory, deposit construction, sampling and seeding, export, decoding, rendering, and evaluation under the simulation’s available resources and field-output schedule. End-to-end measurements on in-situ settings can assess how the fixed-budget reduction framework can be integrated into production simulation workflows.

The temporal experiments establish a useful but limited operating regime. On this series, relocation-enabled warm starts approach independently trained water-fraction encodings only at the tested dense strides of 501--591 cycles; the larger 696--1{,}860-cycle strides remain below their independent counterparts. These cycle counts are properties of this dataset and storage cadence, not a general temporal threshold. The warm-start benefit is also field dependent: it improves water fraction at cycle 17{,}260 but not temperature or asteroid fraction. In addition, the dense-stride curves peak before the final relocation pass and partially decline afterward. This behavior motivates an ablation of relocation timing and duration to characterize post-relocation settling. A longer post-relocation phase or an adaptive decision to skip the final pass are appropriate directions to evaluate. Relocation preserves the prescribed primitive count but deliberately does not preserve primitive correspondence between consecutive models. In the evaluated dense-stride setting, it can improve time-to-quality for independently stored models rather than enabling delta-coded temporal storage. Future work could investigate correspondence-aware reassignment to combine warm starts with efficient temporal differencing.

The current scale and anisotropy constraints restrict the smallest representable features, motivating evaluation of adaptive bounds for thin features and coverage under the same fixed budget. The HACC hash-grid baseline attains the highest reported reconstruction accuracy, but at the largest reported model size and with neural field evaluation rather than an explicit scalar-primitive representation. The Gaussian representation instead provides a fixed primitive count, direct kernel evaluation, and scalar parameters that can be reused under different transfer functions. Future work could evaluate per-dataset W-VEG parameter sweeps; Table~\ref{tab:wveg} reports results obtained with W-VEG's released public configuration and capacity settings used in this study.

Timing measurements are single-run optimization-loop measurements on one laptop GPU, and the 8-bit fidelity result is demonstrated for the reported HACC models rather than established as a universal quantization bound. Broader hardware, dataset, and quantization sweeps remain necessary before generalizing these runtime and storage observations.
\section{Conclusion}
\label{sec:conclusion}

We developed and evaluated a fixed-budget, sample-based Gaussian encoder for scalar scientific fields represented on regular grids, by particles, or by adaptive hexahedral cells. The method initializes primitives from the native input samples and refines them against a field-space reference while preserving a prescribed primitive count. The full-precision representation requires 48 bytes per primitive; the reported 8-bit exports occupy approximately 12 bytes per primitive on disk, including lossless archive compression. This formulation retains an explicit scalar representation that can be decoded independently of a particular transfer function. On controlled regular-grid comparisons with matched primitive budgets, stored sizes, and decoding, the sample-based encoder increased rendered PSNR from 22.8 to 41.1\,dB on Vortex and from 23.7 to 28.9\,dB on Bubble Plume relative to the grid-seeded encoder~\cite{martin2026fixedbudget}. Its 32K-primitive Vortex model reached 49.67\,dB voxel PSNR, exceeding the 44.90\,dB result of the 1.4M-primitive grid-seeded model by 4.77\,dB with approximately $44\times$ fewer primitives. For HACC, the matched $256^3$-supervision control, in which both methods were supervised and evaluated against the same particle-derived reference, improved rendered PSNR from 28.7 to 42.5\,dB at 300K primitives. Under the configured public-release schedule of W-VEG, the sample-based representation also achieved higher voxel PSNR at nearly matched primitive counts on Vortex and Miranda. The Deep Water Impact experiment further showed that primitives can be initialized from 44.9 million adaptive-mesh cells and refined against a deposited reference field, producing a 19.2\,MB encoding with 31.4\,dB rendered PSNR. Warm starts with budget-preserving relocation reduced optimization needed to approach independently trained water-fraction encodings for the dense stored strides evaluated in the temporal series. The results therefore identify a practical operating regime in which temporal reuse is most effective when successive stored fields are sufficiently close for relocation to adapt primitive capacity to evolving structure. Overall, the results establish a single fixed-capacity representation interface across the evaluated grid, particle, cell, and temporal settings while providing predictable storage and strong reconstruction quality per primitive. Direct particle- and cell-query supervision, support for more general cell types, and correspondence-preserving temporal updates are important next steps toward grid-free optimization and efficient temporal differencing.

\bibliographystyle{abbrv-doi}
\bibliography{references}
\end{document}